\newcommand{\bea}{\begin{eqnarray}}
\newcommand{\eea}{\end{eqnarray}}
\documentclass[twocolumn,pra,preprintnumbers,amsmath,amssymb,superscriptaddress]{revtex4}
\usepackage{graphicx}
\newcommand{\p}{\partial}

\def\6#1{{\underline{#1}}}
\def\m6#1{{\underline{#1}\,}}

\newdimen\Tdim
\def\ispan{{\setbox0=\hbox{i}%
\Tdim\ht0\advance\Tdim\dp0\rule[-\dp0]{0pt}{\Tdim}}}
\def\jspan{{\setbox0=\hbox{j}%
\Tdim\ht0\advance\Tdim\dp0\rule[-\dp0]{0pt}{\Tdim}}}
\def\Tspan#1{{\setbox0=\hbox{#1}%
\Tdim\ht0\advance\Tdim\dp0\advance\Tdim.55ex\rule[-\dp0]{0pt}{\Tdim}\box0}}

\def\be{\begin{eqnarray}}
\def\ben{\begin{eqnarray*}}
\def\ee{\end{eqnarray}}
\def\een{\end{eqnarray*}}
\def\Tr{{\rm Tr}}

\def\p{\partial}

\def\=:{=\hspace{-.7em}\raisebox{1.1ex}{.}\hspace{.1em}\raisebox{-0.2ex}{.} }

\newcommand {\beq}{\begin{eqnarray}}
\newcommand {\eeq}{\end{eqnarray}}
\newcommand {\non}{\nonumber\\}

\begin{document}

\preprint{RIKEN-MP-9}
\preprint{RIKEN-TH-204}

\title{
Chiral Magnetic Effect from Q-balls}

\author{Minoru  {\sc Eto}}\author{Koji {\sc Hashimoto}}\author{Hideaki {\sc Iida}}
\affiliation{
{\it Mathematical Physics Lab., RIKEN Nishina Center, Saitama 351-0198,
Japan }}

\author{Akitsugu {\sc Miwa}}\email[]{meto, koji, hiida, amiwa(at)riken.jp}
\affiliation{
{\it Theoretical Physics Lab., RIKEN Nishina Center, Saitama 351-0198,
Japan }}

\begin{abstract}
We apply a generic framework of linear sigma models for revealing a mechanism 
of the mysterious phenomenon, the chiral magnetic effect, in quark-gluon plasma. 
An electric current arises along a background magnetic field, which is given 
rise to by Q-balls (non-topological solitons) of the linear sigma model with axial anomaly. 
We find additional alternating current due to quark mass terms. 
The hadronic Q-balls, baby boson stars, may be created in heavy-ion collisions. 
\end{abstract}

\maketitle


It is widely believed that QCD has a phase transition between the hadronic phase 
and the quark-gluon plasma (QGP) phase at finite temperature and density.
Experimental searches for QGP in relativistic heavy ion collisions have been 
revealed that QGP has highly nontrivial properties, such as its perfect fluidity 
\cite{Huovinen:2001cy,Adler:2003kt,Adams:2003am}, for example.
The chiral magnetic effect (CME) \cite{Kharzeev:2007jp,Fukushima:2008xe} 
is one of the most striking phenomena in QGP which has been recently studied
from the theoretical and experimental viewpoints.
The CME,
the separation of electric charge along the axis of an external electromagnetic fields, 
was predicted as a direct evidence of the (not global but local) strong CP violation
 under very intense external magnetic fields, and was observed in heavy ion collisions.
Recently, an experimental evidence was presented by the STAR Collaboration at RHIC \cite{Voloshin:2009hr}.
Since the discovery of the evidence,
CME has been actively studied using non-perturbative techniques in QCD:
P-NJL model \cite{Fukushima:2010fe}, holographic QCD \cite{Rebhan:2009vc}, 
lattice QCD \cite{Buividovich:2009wi}, and so on.

In this short note, in order to understand CME in QGP,
we consider a generic linear sigma model (LSM) which is widely used 
as a key tool to understand the phase transitions.
We find a universal mechanism for CME which is given rise to by
a stable non-topological solitonic configuration of the (pseudo-)scalar mesons, 
so-called Q-ball \cite{Coleman:1985ki}.
Interior of the Q-ball is in the hadronic phase where the
scalar mesons condense, while it is QGP outside the Q-ball.
We find that the electric current along the external magnetic field arises
in a similar manner discussed in literature. In addition, as a consequence of the
Q-ball,  the electric current not only is a direct current but also has
a small alternating current from quark mass terms.

The generic LSM of the scalar mesons 
$\Phi_{ij} = \bar{q}_{\rm R}^j q_{\rm L}^i$ is
given by \footnote{
In this paper, we do not specify the scalar potential. It can be in general
written as a function of the chiral symmetry invariant operators
$V = \alpha \Tr[\Phi\Phi^\dag] + \beta \left(\Tr[\Phi\Phi^\dag]\right)^2 + 
\gamma \Tr\left[(\Phi\Phi^\dag)^2\right] + \cdots$, where
the coefficients are functions of temperature and density.
} 
\beq
{\cal L}_{\rm eff} 
= 
\Tr\left[ \partial_\mu \Phi \partial^\mu \Phi^\dag - M(\Phi + \Phi^\dag)
\right] - V(\Phi\Phi^\dag)
\nonumber \\
+ A \left(\det \Phi + \det \Phi^\dag\right),
\eeq
where the metric is taken to be 
$\eta_{\mu\nu} = {\rm diag}(+,-,-,-)$.
The matrix $M$ is proportional to the quark mass matrix 
$M \propto {\rm diag}(m_{\rm u},m_{\rm d},m_{\rm s})$,
and the last term is a manifestation of the $U(1)_{\rm A}$
anomaly in QCD.
This Lagrangian enjoys the same symmetries as QCD.
$\Phi$ is singlet under $SU(3)_{\rm C} \times U(1)_{\rm B}$.
The chiral $SU(3)_{\rm L} \times SU(3)_{\rm R}$ symmetry and 
$U(1)_{\rm A}$ acts on $\Phi$ 
as $\Phi \to e^{i\alpha} U_{\rm L} \Phi U_{\rm R}^\dag$, 
which are the exact symmetries if $M=0$ and $A=0$.
The currents corresponding to the axial part 
of these symmetries $\Phi \to e^{i \alpha} U \Phi U$,
are given by 
$J_\mu^{5,a} 
= 
i {\rm Tr}
[T^a
(
\Phi \partial_\mu \Phi^\dag 
+ 
\partial_\mu \Phi^\dag \Phi
-
\Phi^\dag \partial_\mu \Phi
-
\partial_\mu \Phi \Phi^\dag
)]$, 
where $\lambda^a = 2 T^a$ $(a=1,\ldots,8)$ are 
the Gell-Mann matrices for flavor $SU(3)$ 
and $T^0 ={1\over \sqrt{6}}{\boldsymbol 1}_3$.

In order to discuss CME, we consider the electromagnetic field.
The electromagnetic couplings of the quarks in QCD give rise to additional anomalies 
for the diagonal elements of the axial currents, $J_\mu^{5,a=0,3,8}$. From 
the effective theory point of view, 
these anomalies generate couplings
between the diagonal pseudo-scalars and 
the electromagnetic field.
Our idea for the mechanism of CME 
is that non-trivial background for 
the diagonal pseudo-scalars results in 
the electric current through these anomalous couplings.

Motivated by such an idea,
we concentrate on the diagonal pseudo-scalars 
and the overall $\sigma$ field.
Namely, we restrict $\Phi$ as
$\Phi = {\rm diag}(\Phi_1,\Phi_2,\Phi_3) 
= \sigma e^{i \eta' T^0 + i \pi^0 T^3 + i \eta T^8}$.
Then the Lagrangian ${\cal L}_{\rm eff}$ of the 
LSM, before coupled to the electromagnetism, 
is simplified to
\begin{equation}
\sum_i \!
\big(
| \partial_\mu \Phi_i |^2 - M_{ii}(\Phi_i + \Phi_i^\ast)
\big)
- V \big( |\Phi_i|^2 \big) + A 
\Big( \!\prod_i \!  \Phi_i + \!\prod_i \! \Phi_i^\ast \! \Big).
\nonumber
\end{equation}
If $M_{ii}=A=0$, this model respects the 
$U(1)_{{\rm L}-{\rm R}}^3$ part of 
the axial $U(3)_{{\rm L}-{\rm R}}$ symmetry.
The corresponding 
current conservation laws $\partial^\mu J^{5,a}_{\mu} = 0$
($a=0,3,8$) follow from the equation of motion,
where the explicit forms of 
the currents 
are given by
$
J_\mu^{5,a} = 2 i  
\Sigma_i T^a_{ii}
(
\Phi_i \partial_\mu \Phi^\ast_i
- 
\Phi^\ast_i \partial_\mu \Phi_i
)
$.

Next we couple the electromagnetic field to this system.
Since all the scalars we are considering 
are neutral, the only possible couplings
are the anomalous ones explained above.
The explicit forms of such terms can be
found by requiring that they should contribute to 
the anomalous current conservation law correctly.
Our proposal is \footnote{
A similar term has been introduced in a non-linear sigma model in Ref.~\cite{Son:2004tq}.
}
\beq
{\cal L}_{5\!\!\!/} 
&=& \frac{3ie^2}{32\pi^2} F_{\mu\nu}\tilde F^{\mu\nu}\ 
\sum_i q_i^2 (\log \Phi_i - \log \Phi_i^\ast)
\nonumber \\[-2mm]
&=& \frac{3e^2}{16\pi^2}\epsilon^{\mu\nu\rho\sigma}P_\mu A_\nu F_{\rho\sigma},
\eeq
where $\tilde{F}^{\mu\nu} = \frac{1}{2} \epsilon^{\mu\nu\rho\sigma}F_{\rho\sigma}$
($\epsilon^{0123}=1$) and
each $q_i$ is the $i$-th element of the 
electric charge matrix for $({\rm u,d,s})$ 3-flavors, 
${\rm diag}({2 \over 3},-{1 \over 3},-{1 \over 3})$.
We have introduced 
$P_\mu = (P_0, \vec P) \equiv \p_\mu 
{\rm Im}\left(\sum_i q_i^2\log\Phi_i\right)$
and ignored the total derivative term in the right-most hand.
With this term, the anomalous current conservation law 
is derived by using the Euler-Lagrange equation for $\Phi_i$ 
following from ${\cal L}_{\rm eff} + {\cal L}_{5\!\!\!/}$,
\beq
\p_\mu J^{\mu a}_5 & \!\!= \!\!& 
\sum_j T^a_{jj}
\Big(
- \frac{3 e^2}{8\pi^2} q_j^2 F_{\mu\nu} \tilde F^{\mu\nu}
- 2iM_{jj}(\Phi_j-\Phi_j^\ast)
\Big)
\nonumber \\[-2mm]
& &
+ \sqrt{6}iA
\Big(\prod_i \Phi_i  - \prod_i \Phi_i^* \Big) \delta^{a0}.
\label{eq:anm_current}
\eeq
Once we restrict $\Phi$ to the pseudo-scalar neutral mesons 
($\eta'$, $\pi_0$ and $\eta$), this reproduces the standard known form of the anomalous law.

As mentioned above, the additional interaction 
${\cal L}_{\rm 5\!\!\!/}$ 
plays a role of an extraordinary source for 
the electromagnetic field $((-1/2)A_\mu {\cal J}^\mu)$\footnote{
The factor $1/2$ is needed since ${\cal J}^\mu$ itself includes $A_\mu$.
}:
\beq
{\cal J}^\mu = - \frac{3e^2}{4\pi^2}P_\nu  \tilde F^{\nu\mu}.
\label{eq:cme_current}
\eeq
The Maxwell equations
derived from the full Lagrangian including the Maxwell term,
$-{1 \over 4} F_{\mu \nu} F^{\mu \nu} + 
{\cal L}_{\rm eff} + {\cal L}_{\rm 5\!\!\!/}$;
are modified indeed,
\beq
\vec{\nabla} \times \vec B - \frac{\p \vec E}{\p t} = \vec J_{\rm em} + \vec {\cal J},\qquad
\vec\nabla \cdot \vec E =  \rho_{\rm em} + {\cal J}_0.
\label{eq:mxw}
\eeq
with 
$\vec\nabla\times \vec E + \frac{\p \vec B}{\p t} = 0$ and 
$\vec\nabla \cdot \vec B = 0$.
Here $J_{\rm em}^\mu = (\rho_{\rm em},\vec J_{\rm em})$ stands for
the usual electric current, which vanishes for our model with  the
charge-neutral scalars.
Thus the triangle anomaly in QCD 
results in the electromagnetic currents.

Note that these modified Maxwell equations 
are formally identical to those in the Maxwell-Chern-Simons theory
\cite{Kharzeev:2009fn} if the small fluctuation of $\theta$ in \cite{Kharzeev:2009fn}
is replaced by our LSM field $\Phi$.
Furthermore, relation between CME and 
axion strings and domain walls was studied in \cite{Gorsky:2010dr}.
It has been proposed that CME occurs once the $\theta$ 
parameter locally fluctuates \cite{Kharzeev:2007jp,Fukushima:2008xe}.
Since current experiments suggest that 
the $\theta$ parameter in the bare QCD Lagrangian is very small 
$\lesssim 10^{-10}$ \cite{Baker:2006ts},
the origin of such a fluctuation 
is attributed to the effect of the medium in the QGP phase.

In our approach with the generic LSM, on the other hand, CME is triggered by 
the Q-balls \cite{Coleman:1985ki} of the LSM field $\Phi$, 
which are stable finite-size non-topological solitons. 
In the following, we shall show that the current of the CME is given by a typical frequency 
$\omega$ attributed to the Q-ball,
as will be found in (\ref{eq:cme}).
Note that our argument is independent of the locally fluctuating $\theta$ mentioned above.

In order to prevent inessential complexities, 
hereafter we will consider one-flavor model 
\beq
{\cal L} &= - \frac{1}{4} F_{\mu\nu}F^{\mu\nu} \! + \! |\p_\mu \Phi|^2  
\! - \!  V(|\Phi|^2) 
\! + \!  h (\Phi + \Phi^*) \! - \! \frac12   A_\mu {\cal J}^\mu, 
\nonumber
\label{eq:lag_one}
\eeq
where 
${\cal J}^\mu \!=\! - \frac{3e^2}{4\pi^2}
\tilde F^{\nu\mu} \p_\nu
{\rm Im}
\left(
q^2 \log \Phi
\right)
$ and
$h$ includes both the quark mass term $M$ and the anomaly term $A$.
This Lagrangian has $U(1)_{\rm A}$ symmetry if the last two terms vanish.
Let us first construct the Q-ball in $h=0$ limit.
We deal with the electromagnetic field as a background field.\footnote{
Namely, we ignore back reactions from the electromagnetic fields.
Precisely speaking, we take the leading order in the expansion 
with respect to the electric charge $e$.}
The Q-ball's charge, Q-charge, is the axial charge in this one-flavor model.

\begin{figure}[t]
\begin{center}
\includegraphics[height=3cm]{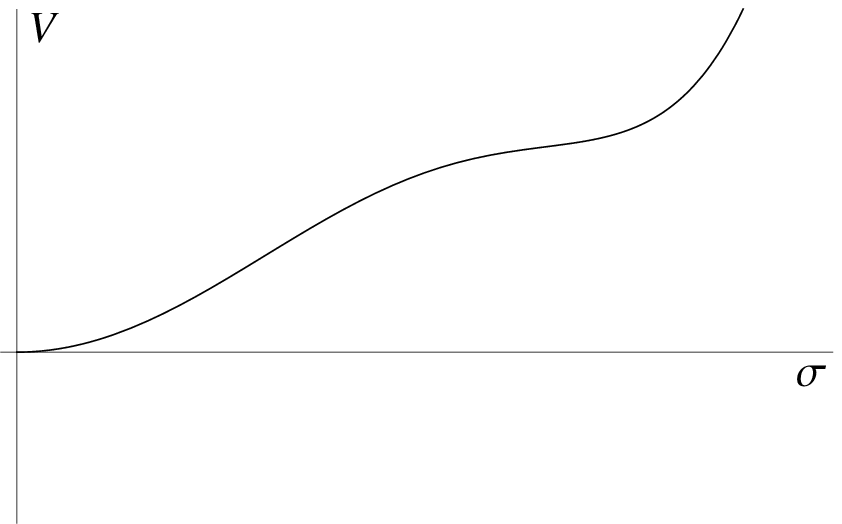}
\caption{{\footnotesize A typical form of the scalar potential $V$. 
There is a true vacuum at $\sigma = 0$ (QGP).}}
\label{fig:potential}
\includegraphics[height=3cm]{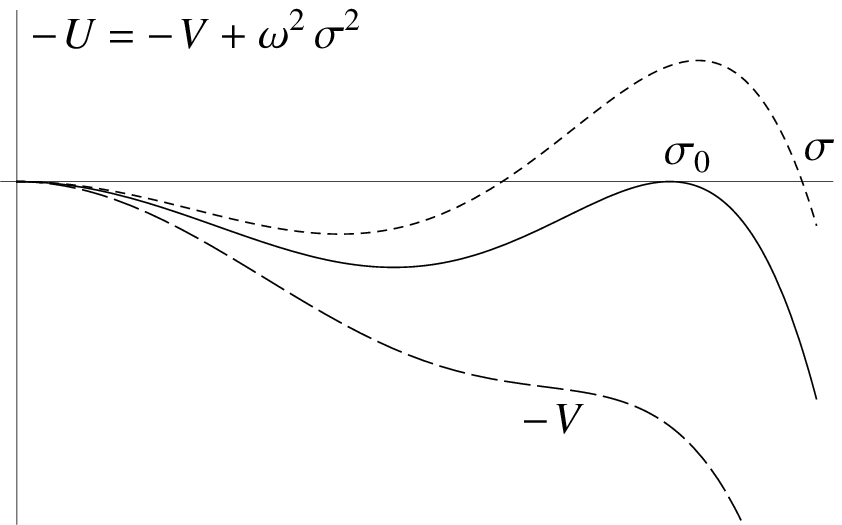}
\caption{{\footnotesize The short-dashed line is a typical form of $U$ which allows 
the Q-ball. The solid line is the special case in which the two extremal
values coincide, $U(0)=U(\sigma_0)=0$.}}
\label{U-potential}
\end{center}
\vspace{-0.8cm}
\end{figure}

The existence of Q-balls does not depend on the details of the system \cite{Coleman:1985ki}. 
One requirement is 
that the scalar potential $V(\sigma^2)$ 
has a true vacuum at $\sigma=0$ (QGP)
as is given in Fig.~\ref{fig:potential}.
Let us make the following ansatz for a spherically symmetric 
$\eta'$-ball
\beq
\Phi = \sigma(r) e^{i\eta'(t)}, \qquad \eta'(t) = \omega t,
\eeq
with $r = |\vec x|$.
The Euler-Lagrange equation for the profile function $\sigma$ leads
\beq
\sigma'' + \frac{2}{r}\sigma' - {1 \over 2} \frac{\p U}{\p \sigma} = 0,\qquad
U(\sigma) \equiv V(\sigma^2) - \omega^2\sigma^2.
\label{eq:EL-sigma}
\eeq
This system can be interpreted as a one dimensional 
classical mechanics with the potential $-U$ 
where $r$ is ``time'' and $\sigma$ is ``position.''
The term $(2/r)\sigma'$ plays a role of the damping force.
Roughly speaking, the Q-ball is 
a solitonic solution connecting two extrema of $-U(\sigma)$.
Therefore, the Q-ball exists when the extrimum at 
$\sigma \neq 0$ appears 
and it is higher than that at $\sigma = 0$, 
as is depicted in Fig.~\ref{U-potential}.
The smoothness and finiteness of the solution requires 
$\sigma'=0$ at both $r=0$ and $r=\infty$.
Coleman showed that a solution exists if $\omega$ is 
in the range $\omega_0^2 < \omega^2 < \mu^2$ 
\cite{Coleman:1985ki}, where $\mu$ specifies the curvature of
the potential $V$  
at $\sigma=0$, $\mu^2=\partial V / \partial \sigma^2 |_{\sigma=0}$, 
and $\omega_0$ is Q-independent frequency,
see below.

\begin{figure}[t]
\begin{center}
\includegraphics[height=4cm]{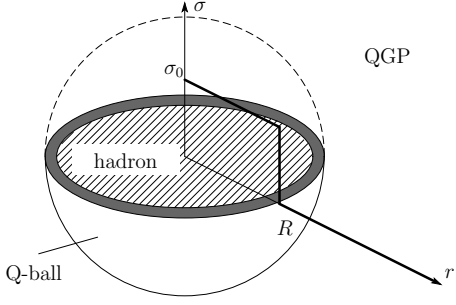}
\caption{{\footnotesize 
A schematic picture of the spherical Q-ball.}}
\label{fig:qball}
\end{center}
\vspace{-0.8cm}
\end{figure}

The Q-ball can be best understood in the 
large Q-charge limit, where
$\sigma(r)$ resembles a smoothed-out step function.
Then we assume that 
for small $r$ less than a certain radius $R$, 
$\sigma = {\rm const.} > 0$ whereas for large $r > R$, 
$\sigma = 0$, see Fig.~\ref{fig:qball}. Namely,   
\beq
\Phi\big|_{r<R} = \sigma e^{i\omega t},\quad
\Phi\big|_{r>R} = 0.
\label{eq:large_q}
\eeq
We ignore the contributions from the transition zone around $r=R$ 
(surface of the Q-ball)
which may be subdominant compared with the volume ones.
In the limit, we obtain the energy 
$E = \omega^2 \sigma^2 {\cal V} + V(\sigma^2) {\cal V}$ 
and Q-charge $Q = 2\omega \sigma^2 {\cal V}$
with ${\cal V}$ being the volume of Q-ball, 
${\cal V}= 4\pi R^3/3$. 
A stable solution with fixed Q-charge is
given by minimizing $E$ with respect to 
three variables $(\sigma, \omega, {\cal V})$ with 
the constraint $Q$ fixed.
First, $E$ is expressed in terms of $(\sigma,{\cal V})$ as 
$E = {1\over 4} {Q^2 \over \sigma^2 {\cal V}} + V(\sigma^2){\cal V}$.
Then by minimizing it with respect to ${\cal V}$
one gets
$E = Q \sqrt{\frac{V}{\sigma^2}}$ and 
${\cal V} = \frac{Q}{\sqrt{4 \sigma^2 V}}$,
which also determines $\omega$ in terms of $\sigma$
as $\omega^2 \sigma^2 = V(\sigma^2)$.
Finally, we determine the value of $\sigma$
by minimizing $E$ with respect to $\sigma$.
Let $\sigma_0 \neq 0$ be the value of $\sigma$
for which $E$ takes its minimum $E_0$,
and $\omega_0$ be the corresponding frequency.
Then, in summary, these values are determined by 
\beq
E_0 = Q \omega_0, \qquad 
\omega_0^2 
= V'(\sigma_0^2), 
\qquad \omega_0^2 \sigma_0^2 = V(\sigma_0^2),
\label{eq:Q-ball(Ews)}
\eeq
where the prime stands for $V'(\sigma^2) = \p V(\sigma^2)/\p(\sigma^2)$.
The last two equations determine both $\sigma_0$ and $\omega_0$
independently from $Q$.
In fact these equations mean that the two extremal values of 
$U(\sigma) = V(\sigma^2) - \omega_0^2 \sigma^2$ coincide
and $\sigma_0$ is one of the extrema with $\sigma_0 \neq 0$, see 
Fig.~\ref{U-potential}.
In this case, since $\sigma(r)$ can spend arbitrary long ``time'' $r$
at the extremum $\sigma = \sigma_0$, the solution can have 
an arbitrary large volume $\cal V$, and hence also the large Q-charge.
Also, since the damping force in \eqref{eq:EL-sigma} is 
negligible after the long time, the profile $\sigma(r)$
is well approximated by the smoothed-out step function.

With the $Q$-ball at hand, we now see from Eq.~(\ref{eq:cme_current}) 
that the electric current arises along a constant background magnetic 
field $\vec B$ (the electric field is assumed to be zero)
\beq
\vec {\cal J} =  \frac{3e^2}{4\pi^2} q^2 \omega \vec B.
\label{eq:cme}
\eeq
Here, we see that {\it CME is a consequence of the existence of the Q-ball.}
The magnitude of our CME current is given dynamically by $\omega$ 
of the Q-ball. 
As found in (\ref{eq:Q-ball(Ews)}), 
$\omega_0$ is given by the LSM potential characterized typically by
$\Lambda_{\rm QCD}$. So it is natural to assume $\omega_0\sim\Lambda_{\rm QCD}$.
Using the expected value of the magnetic field 
$eB\sim 10^4$ [MeV${^2}$] in heavy ion collisions 
\cite{Kharzeev:2007jp},  we obtain the magnitude of the CME current as 
$|\vec{J}|\sim 10^5$ [MeV$^3$] $\sim 10^{-2}$ [fm$^{-3}$]. 

Also, natural size of the Q-ball is $\sim Q^{1/3}{\rm [fm]}$, 
if all the dimensionful parameters are of order $\Lambda_{\rm QCD} \sim 1 [{\rm fm}^{-1}]$.

Note that this important frequency $\omega$ plays a role of 
the so-called chiral chemical potential
$\mu_5$ \cite{Fukushima:2008xe}.
The chemical potential can be introduced 
through the change
$\partial_0 \Phi \to (\partial_0 + i \mu_5)\Phi$ in \eqref{eq:lag_one}.
The relevant terms for this change 
are the ones with time derivatives:
\beq
&& |\partial_0 \Phi|^2
-
\partial_0 {\rm Im} (\log \Phi) 
{3 e^2 q^2\over 8 \pi^2} 
\vec A \cdot \vec B
\nonumber \\
&& \to 
|(\partial_0 \! +\! i \mu_5)\Phi|^2
-
\big(\partial_0 {\rm Im} (\log \Phi)\! +\! \mu_5\big)
{3 e^2 q^2\over 8 \pi^2} 
\vec A \cdot \vec B.\qquad
\label{chemical}
\eeq
Taking the Q-ball solution $\Phi= \sigma(r) e^{i\omega t}$ 
in the theory without $\mu_5$ is equivalent to 
considering a static solution 
$\Phi = \sigma(r)$ in the theory with $\mu_5$,
if we identify $\mu_5 = \omega$.
In the latter case, the current \eqref{eq:cme}
is supplied by the last $\mu_5 \vec A \cdot \vec B$ 
term in \eqref{chemical}.

Since the $Q$-charge is preserved, this non-topological soliton 
is fairly stable. However, the $h(\Phi+\Phi^*)$ term in Eq.~(\ref{eq:lag_one}) 
breaking explicitly $U(1)_{\rm A}$ may
destabilize the Q-ball, which would result in destroying a constant supply of the 
electric current. So,
let us next analyze the effect of $h(\Phi + \Phi^*)$.
We expect that, if $h$ is sufficiently small, $U(1)_{\rm A}$ is broken only weakly and
Q-ball still lives long. In the following, we shall derive a condition for $h$ to have the 
stability, and find that Q-ball are fairly stable, but with a new interesting feature of
alternating CME current component.

We treat $h$ as a small parameter and we expand
fields with respect to a small dimensionless parameter $\epsilon$ as
\beq
\Phi = \Phi_0 + \epsilon \Phi_1 + \epsilon^2 \Phi_2 + \cdots,\qquad
\epsilon = \frac{h}{\sigma_0\omega_0^2},
\eeq
where $\Phi_0$ is the $Q$-ball solution in $\epsilon\to0$ limit.
Again, we consider the large Q-ball limit given in Eq.~(\ref{eq:large_q}).
We would like to solve the equations of motion 
\beq
F = \p_\mu \p^\mu \Phi + \Phi V' - h = 0,
\eeq
order by order in $\epsilon$
as $F = F_0 + \epsilon F_1 + \epsilon^2 F_2 + \cdots = 0$. 
The zeroth order $F_0 = 0$ is the Q-ball equation which we have solved.
As we have seen, this gives us $\omega_0^2 = V'(\sigma_0^2)$.
The next-to-leading order is $F_1=0$ with $F_1$ given by 
\beq
\p_\mu \p^\mu \Phi_1 \! + \!  \Phi_1 \! V'(|\Phi_0|^2) 
\! + \!  \Phi_0  \! \left(\Phi_0^*\Phi_1 \! + \!  \Phi_0\Phi_1^*\right)
\! V''(|\Phi_0|^2) \! - \! 
 \sigma_0 \omega_0^2.
\nonumber
\eeq
We solve this equation in two region, $r < R$ and $r>R$, separately.
In the former region we put $\Phi_0 = \sigma_0 e^{i\omega_0t}$ while $\Phi_0 = 0$ in the latter region. 
Therefore, the next-to-leading order solution is given by
\beq
\Phi_1\big|_{r<R} & = & 
\frac{3\alpha_1 - 1}{3\alpha_1 + 2 } \sigma_0+
\frac{1}{3\alpha_1 + 2}\sigma_0e^{2i\omega_0t}, 
\eeq
and $\Phi_1\big|_{r>R} =\sigma_0\omega_0^2/V'(0)$, 
with $\alpha_1 \equiv \frac{V'(\sigma_0^2)}{V''(\sigma_0^2)\sigma_0^2}$.
The next-to-next-to-leading order is readiliy solved by
\beq
& & \Phi_2\big|_{r<R} 
= \frac{
-36\left(2 + \alpha_2^{-1} \right)\alpha_1^2 + 9 \alpha_1 + 2
}{8\left(3\alpha_1+2\right)^2} \sigma_0e^{-i\omega_0t} \non
\nonumber\\
& &
- \frac{9(2\!+\!\alpha_2^{-1}) \alpha_1^2\!-\! 6 \alpha_1\!+\!2}{
2(3\alpha_1+2)^2}\sigma_0 e^{i\omega_0t}
\!+\! \frac{3(-3\alpha_1\!+\!2)}{8(3\alpha_1+2)^2}\sigma_0 e^{3i\omega_0t},
\nonumber 
\eeq
and $\Phi_2\big|_{r>R} = 0$, where we have defined
$\alpha_2 = \frac{V''(\sigma_0^2)}{V'''(\sigma_0^2)\sigma_0^2}$.
The mass of the Q-ball up to this order is evaluated as
\beq
E 
= Q \omega_0 \!+\! {1 \over 2}
\left[
\frac{3\left(-3(5\!+\!2\alpha^{-1}_2)\alpha_1^2 \!+\! 3 \alpha_1 \!+\! 2 \right)}{
(3\alpha_1+2)^2}
+
{\omega_0^2 \over \mu^2}
\right]  Q\omega_0  \epsilon^2 
\nonumber
\eeq
where we have shifted the origin of the energy
in such a way that the energy density outside 
the Q-ball becomes zero, and 
$Q=2 \omega_0 \sigma_0^2 {\cal V}$ as before. 
Note that the first contribution starts at the order of 
$\epsilon^2$ and the next is of order $\epsilon^4$. 
Therefore, the variation in energy 
is negligible if $\epsilon$ is sufficiently small,
namely $h \ll \sigma_0\omega_0^2$.
By using $\omega_0^2 \sigma_0^2 = V(\sigma_0^2)$
(see \eqref{eq:Q-ball(Ews)}),
this condition can be rewritten as 
$h \sigma_0 \ll V(\sigma_0^2)$.
This condition is quite natural since it just
means that the perturbation $h(\Phi + \Phi^\ast)$ is 
small compared to the original potential $V(|\Phi|^2)$.
Hence we can 
expect that the
Q-ball is stable against the perturbation.

Interestingly, a contribution of order $\epsilon$
arises in the electric current as
\beq
\vec{\cal J} =  \frac{3e^2}{4\pi^2} q^2 \vec B \omega_0 \left(1 -
\epsilon \frac{3\alpha_1\!-\!2}{3\alpha_1\!+\!2} \cos \omega_0 t + \cdots
\right).
\eeq
Thus the quark mass term and the $U(1)_{\rm A}$ anomaly in QCD
eventually give rise to a small alternating CME current.
This is a new feature of CME by the Q-ball.

Let us finally make a comment on a possibility of the Q-balls by
the other pseudo-scalar mesons such as $\pi^0$ and $\eta$.
In addition to $U(1)_{\rm A}$, there exists an axial part of
the chiral symmetry $SU(3)_{\rm L-R} \in SU(3)_{\rm L} \times 
SU(3)_{\rm R}$ in QCD.
It is straightforward to construct the Q-balls by using a 
$U(1)^2_{{\rm L}-{\rm R}}$ subgroup
in $SU(3)_{\rm L-R}$, for instance, a pionic Q-ball with the $T^3$
generator. 
Since the $U(1)_{\rm L-R}^2$ is also anomalous 
by the electromagnetic interaction,
the CME current arises as in the case of the $\eta'$-ball \footnote{
Coleman \cite{Coleman:1985ki} mentioned that it is an open question whether 
a gauged $U(1)$ symmetry allows a Q-ball or not, which reflects in our case
with a question of having the CME with
Q-balls of charged mesons, such as $\pi^{\pm}$. }.

In summary, we present a useful formalism of LSM which can explain CME in QGP
via the non-topological soliton, Q-balls.
The electric current arises along the external magnetic field and it has a small
alternating current as a consequence of quark mass terms and
the $U(1)_{\rm A}$ anomaly in addition.
Furthermore, the interior of the Q-ball is the hadronic phase.
So we predict that there may be a lump of hadrons, Q-ball, in QGP.
It might have a certain contribution in the cooling process of QGP and hadronization.

In cosmology, the Q-ball is thought of as a candidate of so-called boson stars \cite{Jetzer:1991jr}.
We hope that our study may open a new direction to create baby boson stars at RHIC, LHC and FAIR.
\\\\
\noindent
{\bf Acknowledgment}: 
The authors would like to thank Koichi Yazaki, Kenji Fukushima and Naoki Yamamoto
for useful comments and
discussions. 
The authors thank the Yukawa Institute for Theoretical Physics at Kyoto University. 
Discussions during the YITP workshops YITP-W-10-02 and YITP-W-10-08 
were useful to complete this work.
The work of M.E. and A.M. is supported by Special Postdoctoral Researchers
Program at RIKEN.  K.H.~is partly supported by
the Japan Ministry of Education, Culture, Sports, Science and
Technology.



\begin{thebibliography}{99}

\bibitem{Huovinen:2001cy}
  P.~Huovinen, P.~F.~Kolb, U.~W.~Heinz, P.~V.~Ruuskanen and S.~A.~Voloshin,
  Phys.\ Lett.\  B {\bf 503}, 58 (2001).


\bibitem{Adler:2003kt}
  S.~S.~Adler {\it et al.}  [PHENIX Collaboration],
  Phys.\ Rev.\ Lett.\  {\bf 91}, 182301 (2003).

\bibitem{Adams:2003am}
  J.~Adams {\it et al.}  [STAR Collaboration],
  Phys.\ Rev.\ Lett.\  {\bf 92}, 052302 (2004).


\bibitem{Kharzeev:2007jp}
  D.~E.~Kharzeev, L.~D.~McLerran and H.~J.~Warringa,
  Nucl.\ Phys.\  A {\bf 803}, 227 (2008).

\bibitem{Fukushima:2008xe}
  K.~Fukushima, D.~E.~Kharzeev and H.~J.~Warringa,
  Phys.\ Rev.\  D {\bf 78}, 074033 (2008).


\bibitem{Voloshin:2009hr}
  S.~A.~Voloshin  [STAR Collaboration],
  Nucl.\ Phys.\  A {\bf 830}, 377C (2009).


\bibitem{Fukushima:2010fe}
  K.~Fukushima, M.~Ruggieri and R.~Gatto,
  Phys.\ Rev.\  D {\bf 81}, 114031 (2010).

\bibitem{Rebhan:2009vc}
  A.~Rebhan, A.~Schmitt and S.~A.~Stricker,
  JHEP {\bf 1001}, 026 (2010).

\bibitem{Buividovich:2009wi}
  P.~V.~Buividovich, M.~N.~Chernodub, E.~V.~Luschevskaya and M.~I.~Polikarpov,
  Phys.\ Rev.\  D {\bf 80}, 054503 (2009).

\bibitem{Coleman:1985ki}
  S.~R.~Coleman,
  Nucl.\ Phys.\  B {\bf 262}, 263 (1985)
  [Erratum-ibid.\  B {\bf 269}, 744 (1986)].

\bibitem{Son:2004tq}
  D.~T.~Son and A.~R.~Zhitnitsky,
  Phys.\ Rev.\  D {\bf 70}, 074018 (2004).



\bibitem{Kharzeev:2009fn}
  D.~E.~Kharzeev,
  Annals Phys.\  {\bf 325}, 205 (2010).

\bibitem{Gorsky:2010dr}
  A.~Gorsky and M.~B.~Voloshin,
  Phys.\ Rev.\  D {\bf 82}, 086008 (2010).

\bibitem{Baker:2006ts}
  C.~A.~Baker {\it et al.},
  Phys.\ Rev.\ Lett.\  {\bf 97}, 131801 (2006).

\bibitem{Jetzer:1991jr}
  For a review, P.~Jetzer,
  Phys.\ Rept.\  {\bf 220}, 163-227 (1992).

\end{thebibliography}
\end{document}